\documentclass[12pt]{article}

\begin{document}

\hyphenation{Min-kow-ski} \hyphenation{cosmo-logical} %
\hyphenation{holo-graphy} \hyphenation{super-symmetry} %
\hyphenation{super-symmetric}

\centerline{\Large \bf }
\vskip0.25cm 
\centerline{\Large \bf }
\vskip0.25cm %
\centerline{\Large \bf Gravitationally dressed Fermi Liquids,}
\vskip0.25cm
\centerline{\Large \bf Quasiunparticles and High $T_c$ Superconductivity}
\vskip0.25cm 
\vskip 1cm

\renewcommand{\thefootnote}{\fnsymbol{footnote}} 
\centerline{{\bf Djordje Minic\footnote{dminic@vt.edu} and J. J.
Heremans\footnote{heremans@vt.edu}}} 
\vskip .5cm 
\centerline{\it Department of Physics, Virginia Tech} 
\centerline{\it Blacksburg, VA 24061, U.S.A.}

\begin{abstract}
We clarify the new concept of gravitationally dressed Fermi liquids 
we have proposed to describe the normal state of high $T_c$ superconductors.
In this note we distinguish between weakly gravitationally dressed Fermi liquids
which fall in the class of the canonical Fermi liquid theory (with quasiparticle excitations), and strongly
gravitationally dressed Fermi liquids (with quasiunparticle excitations) which represent the high dimensional
generalization of Luttinger liquids, with specific features.
Such weakly or strongly gravitationally dressed Fermi liquids lead naturally to the previously proposed effective (weak or strong) gravitationally dressed Landau-Ginsburg description of 
either ordinary or high $T_c$ superconductors. 
\end{abstract}

\vskip .5cm

\setcounter{footnote}{0} \renewcommand{\thefootnote}{\arabic{footnote}}

\newpage

\section{Introduction}

The problem of high temperature superconductivity is one of the most
outstanding puzzles in contemporary physics \cite{htcreview}.
Motivated by the work
on induced gravitational physics in various condensed
matter systems (see for example \cite{abh}, \cite{volovik} for reviews) we have proposed an effective gravitational
Landau-Ginsburg description of high $T_c$ superconductors \cite{us}.
Our approach
concentrated on the subsection of the phase diagram containing the strange
metal/superconducting phase, and thus we were motivated to ask how the usual canonical Wilsonian
thinking associated with naturalness of Fermi liquids should be modified
to lead to a non-Fermi liquid behavior. 
We proposed that for systems with complex Fermi surfaces a concept of
gravitationally dressed Fermi liquids is natural from the point of view of the renormalization group. 

In this note we clarify this proposal for the normal
(``strange metal'') state of the high $T_c$ superconductors
\cite{us}. In particular we distinguish between weakly gravitationally dressed Fermi liquids
which fall in the class of the canonical Fermi liquid theory, and strongly
gravitationally dressed Fermi liquids which represent the high dimensional
generalization of Luttinger liquids, with specific features.
The excitations of Landau Fermi liquids \cite{landau} are particle-like ``quasiparticles'', and
the relevant two-point function has a single particle pole. The corresponding
excitations of gravitationally dressed Fermi liquids are ``unparticle''-like, following
the terminology used in the recent theoretical high energy physics literature \cite{georgi}.
The relevant two-point function does not have a single-particle pole, and 
following the suggestion of \cite{semiholo} we call such excitations ``quasiunparticles''.
The (weakly or strongly) gravitationally dressed Fermi liquids lead naturally to the previously proposed \cite{us}
effective (weakly or strongly) gravitationally dressed Landau-Ginsburg description of 
(ordinary or high $T_c$) superconductors.
Note that our proposal should not be confused with the gravitational or
holographic duals of non-Fermi liquids which are insightfully summarized in \cite{semiholo}.

\section{Fermi liquids vs. non-Fermi liquids}

As noticed in many publications, various condensed matter systems, such as
superfluids or Bose-Einstein condensates, exhibit physical phenomena that
can be interpreted by invoking effective gravity \cite{abh}. In these
contexts, which can be viewed as analog models of gravity, one speaks for
example of an effective acoustic metric, acoustic black holes (i.e. ``dumb
holes'') or emergent relativity.

Given the intricacies of the physics of high temperature superconducting
materials and the complexity of their phase diagram \cite{htcreview}, one might ask whether
the strong electron correlations responsible for these phenomena can induce
effective gravitational effects, thus opening a possibility for a geometric
explanation of some of the outstanding puzzles. In particular, the normal
state of high $T_c$ superconductors should reflect
such emergent gravitational physics associated with the geometry and fluctuations of
the Fermi surface.

Emergent gravity in systems endowed with a complex Fermi surface forces a revisiting of the
Wilsonian approach to the Fermi surface, to ascertain consequences for Fermi
liquid theory in the renormalization group approach, and for the emergence
of non-Fermi liquid behavior \cite{nfl}. From the effective field theory standpoint,
the question can be formulated: how does the low energy Wilsonian action
with effective gravity modify the usual scalings of the effective low energy
field theory of the Fermi surface? We claim that the effective gravity of the 
fluctuating Fermi surface leads to a robust extension of the canonical picture:
a gravitationally dressed Fermi liquid.

\subsection{The Fermi liquid}

To set the stage for our proposal \cite{us} let us summarize the classic effective
field theory of Landau Fermi liquids \cite{joep, flbooks} as succinctly presented by
Polchinski \cite{joep}. One starts with a \textit{natural} (i.e. not-finely tuned) Fermi surface and
decomposes the momenta into the Fermi momentum and a component orthogonal to
the Fermi surface 
\begin{equation}
\vec{p} = \vec{k} + \vec{l},
\end{equation}
and then one considers scaling of energy and momentum towards the Fermi
surface, in other words:
\begin{equation}
E \to sE, \quad \vec{k} \to \vec{k}, \quad \vec{l} \to s \vec{l}.
\end{equation}
The lowest order action, to quadratic order, is then given as 
\begin{equation}
S_{FL} = \int L_{FL} (\psi, \partial \psi) \equiv \int dt d^3 \vec{p} [ i
\bar{\psi} (\vec{p}) \partial_t \psi (\vec{p})- (E(\vec{p}) -E_F(\vec{p}%
))\bar{\psi} (\vec{p}) \psi (\vec{p})].
\end{equation}
Close to the Fermi surface 
\begin{equation}
E(\vec{p}) -E_F(\vec{p}) \sim l v_F, \quad v_F= \partial_{\vec{p}} E,
\end{equation}
so that after the agreed renormalization orthogonal to the Fermi surface
(note that also $t \to s^{-1} t$) one obtains, in both $2+1$ and $3+1$ space-time dimensions: 
\begin{equation}
\psi \to s^{-1/2} \psi.
\end{equation}
This scaling leads immediately to the usual two point function for
a free quasiparticle with a single particle pole
$\frac{1}{\omega + E_F - E(p)} $ (which corresponds to a quasiparticle excitation). In this argument it is crucial
that the transverse part of the momentum is inert. The effectively one-dimensional
scaling is set by the momentum orthogonal to the Fermi surface. Now, by considering
4-Fermi interactions one can see that for generic momenta the 4-Fermi
interaction scales as a \textit{positive} power of $s$ and is thus
irrelevant at low energy. The measure over time contributes one negative
power, the measure over the momenta orthogonal to the Fermi surface
contributes 4 powers and the 4-Fermi interaction contributes $4/2$ negative
powers. The delta function over the 4 momenta generically does not scale. So
the overall generic scaling for the 4-Fermi vertex is indeed 
\begin{equation}
s^{-1+4-4/2} = s^{1}.
\end{equation}
This is valid, except if the momenta are \textit{paired}. In that case the
scaling goes as $s^{0}$, because now the delta function depends only on the
sum of momenta orthogonal to the Fermi surface and due to 
\begin{equation}
\delta (s l) \to s^{-1} \delta(l),
\end{equation}
the 4-Fermi interaction indeed scales as $s^0$ and is marginal. This
encapsulates the usual Cooper pairing phenomenon. Note that this scaling 
is true both in $2+1$ and $3+1$ dimensions, with the $2+1$ dimensional 
case of interest for the layered anisotropic compounds. 
The case of $1+1$ dimensions is special, because 
then the Fermi surface consists of two points, and hence the 4-Fermi 
interaction is automatically marginally relevant.  This special kinematics is responsible for 
the non-Fermi liquid properties of the Luttinger liquid \cite{pw}.

\section{ The gravitationally dressed Fermi Liquid}

In what follows we discuss how the dynamics and fluctuations of the Fermi surface
can substantially change the above canonical reasoning. How could effective gravity arise from the dynamics of the Fermi surface? First, we note that experiments indicate a highly irregular (complex)
Fermi surface in the normal state of high $T_c$ superconductors, resulting from the microscopic
physics \cite{htcreview}. This irregularity in turn could lead to an effective
gravitational description. We claim that this effective gravity is {\it natural} for
a complex Fermi surface. As a concrete model, let us view the Fermi
surface as an incompressible fluid in momentum space. (A hydrodynamic approach to
Fermi liquid theory is presented in \cite{wen}.) In analogy with
the discussion of induced gravity in fluid dynamics (\cite{unruh}, \cite{abh}%
) we can envision generating an effective metric, which in the case of real
irrotational fluids is precisely the acoustic metric \cite{unruh}. This
effective metric is generated from the fluctuations of the fluid density $%
\rho $ and the velocity potential $\phi $ (where the velocity $\vec{v}%
=\nabla \phi $). The underlying space-time action of the moving fluid is 
\begin{equation}
S=\int dx^{4}[\rho \dot{\phi}+\frac{1}{2}\rho (\nabla \phi )^{2}+U(\rho )],
\end{equation}%
with $U(\rho )$ denoting the effective potential. Note that the signs 
in $S$ are consistently defined so that upon the variation of this 
action one obtains the equations of motion for $\rho $ and $\phi $ (the
Euler continuity equation and the Bernoulli energy balance equation) \cite{abh}. 
Note also that the pressure is the negative of the action density in the expression for $S$. 
When these equations of motion are perturbed around the equilibrium values $\rho
_{0}$ and $\phi _{0}$ one is led to the equations for the fluctuations of
the velocity potential $\varphi $. In particular, the equation for the
fluctuations of the velocity potential can be written in a geometric form
\cite{unruh}:
\begin{equation}
\frac{1}{\sqrt{-g}}\partial _{a}(\sqrt{-g}g^{ab}\partial _{b}\varphi )=0.
\end{equation}%
The effective space-time metric has the canonical ADM form of \cite{unruh}
and \cite{abh} (apart from a conformal factor) and has the Lorentzian
signature 
\begin{equation}
ds^{2}=\frac{\rho _{0}}{c}[c^{2}dt^{2}-\delta
_{ij}(dx^{i}-v^{i}dt)(dx^{j}-v^{j}dt)],
\end{equation}%
where $c$ is the relevant effective sound velocity and $v^{i}$ are the
components of the fluid's velocity vector.

Another way to argue for an emergent gravitational background
in the physics of a generically complex Fermi surface, is to invoke the
orthogonality catastrophe (for a nice review consult \cite{wen}).
Orthogonality catastrophe is caused by the vanishing overlap
between the respective many-body wave functions of the deformed and
the undeformed Fermi surfaces. Given the general geometric
structure of quantum theory, it is well known that the overlaps of the many-body
wave functions can lead to induced gauge fields \cite{gphase}, via the connections
in the projective Hilbert space of quantum theory \cite{gqt}.
However, the metric structure is also encoded in the overlaps \cite{gphase, gqt}
and thus an emergent gravitational background could be expected on general
grounds. This emergent gravitational background then generically leads to the gravitational
dressing of the excitations of the Fermi surface.

\subsection{Weakly vs. strongly gravitationally dressed Fermi liquids}

The collective dynamics of the Fermi surface can be considered as
a \textquotedblleft bosonization\textquotedblright\ of the Fermi liquid \cite%
{boson} and in that approach the quasiparticle excitations can be
represented as collective modes of the \textquotedblleft
bosonized\textquotedblright\ Fermi liquid. That approach runs into trouble
with the essential difference between fermions and bosons in spatial
dimensions one (where the bosonization can be used because of the very
special kinematics) and spatial dimensions above one, where such efforts are
largely prohibitive \cite{boson}. 
The same kinematical reasons make it difficult to extrapolate
the Luttinger liquid behavior to higher dimensions \cite{pw}.

In contrast, in our proposal the non-Fermi liquid
behavior originates from the gravitational dressing, itself caused by the
non-trivial geometry and topology of the Fermi surface, an experimental
fact. The weak or strong dressing would correspond respectively to the weak or
strong coupling between the
induced gravity and the collective excitations of
the Fermi surface. The new quasiparticles are the usual fermions, 
propagating, however, in a non-trivial gravitational background and hence dressed by
the gravitational fluctuations. In other words, the collective motion of the
Fermi surface is of an effective gravitational kind (i.e. not spin 0 but
spin 2) and the usual fermionic quasiparticles are now coupled to this
collective spin 2 bosonic mode. Thus we propose that the effective theory of
the strange normal state of high $T_{C}$ superconductors is a
gravitationally dressed Fermi liquid, i.e. the usual Fermi liquid albeit
coupled to gravity:
\begin{equation}
S_{GFL}=\int d^{D}x\sqrt{-g}L_{FL}(\psi ,\nabla \psi ) = \int d^D{x} L_{GFL}.
\end{equation}
The crucial $g$ dependent factors ($\sqrt{-g}$, $\nabla$) arise from the usually inert transverse part of
the momentum. This general discussion of the gravitationally dressed Fermi liquid applies
both to $D=2+1$ (relevant for the the Cu-O or Fe-As 
planes) and $D=3+1$.  

Applying this approach to the Fermi surface in momentum space, by relying on
the induced effective diffeomorphism invariance, we are led to conclude that
the effective action for the fermionic quasiparticle around the Fermi
surface should be a gravitationally dressed action, in which the canonical
scaling dimensions discussed above can be changed by adding gravitational
dressing. In general this would mean that instead of $\psi \to s^{-1/2} \psi$
as discussed above, we should have 
\begin{equation}
\psi \to s^{-1/2 +\alpha} \psi,
\end{equation}
where $\alpha$ denotes gravitational dressing. Such dressing can be explicitly 
computed in simple cases such as the coupling of $1+1$-dimensional gravity 
to $1+1$-dimensional matter \cite{kpz, kkp, kawai, bilal}. Note that this is relevant for the present discussion,
because the scaling towards the Fermi surface, although not the kinematics, is effectively one-dimensional. Thus the
fermionic two point function can be changed to scale as $s^{-1 +2\alpha}$,
indicative of a non-Fermi liquid behavior.

Let us briefly summarize some generic features of a gravitationally dressed
free fermion theory, which defines our proposal more precisely.
A nice presentation of the relevant formalism is given
in \cite{kkp, kawai, bilal} which we follow. The central results for the gravitational
dressing of $1+1$ theories were originally obtained in \cite{kpz}. 

Using the main idea of our proposal which stipulates the
dynamical role of the transverse momenta in the 
canonical treatment of Fermi Liquid theory
and following
\cite{kawai, bilal}, the gravitationally dressed Fermi
liquid theory (GFL) is described by the following Lagrangian 
(in momentum space)
\begin{equation}
L_{GFL} = e (\bar{\psi} i \gamma^a  e_a^{t} \partial_{t}\psi - (E-E_F) Z_E (g) \bar{\psi} \psi),
\end{equation}
where, apart from the gamma matrix $\gamma^a$, we have used the appropriate ``bein'' $e^a_{\mu}$ \cite{kawai, bilal}
($\mu$ being the general space-time index), its inverse $e_a^{\mu}$ and its determinant $e$.
This is necessary for describing the coupling between effective gravity ($g$) of
the Fermi surface and the fermion $\psi$ field.
Note that the effective wave function renormalization $Z_E (g)$ (which is central in
Fermi liquid theory because it defines the residue of the single particle pole \cite{joep, flbooks})
depends on the
effective gravity ($g$) and defines the anomalous dimension of
the effective mass term $(E-E_F) \bar{\psi} \psi$ \cite{kawai}:
\begin{equation}
\Lambda \frac{d E}{d \Lambda} = E \frac{d \log{Z_E}}{d \log{\Lambda}}.
\end{equation}

The crucial point is that, as shown in the classic paper \cite{kpz},
the anomalous dimension of a given operator in $D=2$ Euclidean space-time
(in our particular case, a given fermionic operator
in Fermi liquid theory)
with the canonical scaling dimension $\Delta$, becomes 
a gravitationally dressed scaling dimension $\Delta_g$ determined by
the famous KPZ formula \cite{kpz}:
\begin{equation}
\Delta_g - \Delta = \frac{ \Delta_g(\Delta_g -1)}{\beta}.
\end{equation}
Here $\beta$ is governed by the total central charge
$c$ of the matter theory (in our case, Fermi liquid theory) coupled to gravity
\cite{kpz, kkp, kawai, bilal}:
\begin{equation}
\beta= \frac{1}{12}[ c-13 - \sqrt{(c-1)(c-25)}].
\end{equation}
These classic expressions can be extrapolated using the $\epsilon$-expansion to
$D=2+\epsilon$ Euclidean space-time dimensions \cite{kawai}, which is more
appropriate in general.

The fermionic two-point correlation functions 
become dressed so that instead of
a single particle pole (corresponding to a quasiparticle excitation) we get \cite{bilal} a power law scaling characteristic
of a non-Fermi liquid behavior (corresponding to a quasiunparticle excitation):
\begin{equation}
\frac{1}{\omega + E_F - E(p)} \to \frac{1}{(\omega + E_F - E(p))^{2\alpha +1}}.
\end{equation}
Thus gravitationally dressed Fermi liquids emulate the behavior of
Luttinger liquids even though the physics underlying the
gravitationally dressed Fermi liquids is not constrained by the kinematics as
is the case with Luttinger liquids. Note that given our general construction,
we should not expect the spin-charge separation, in contradistinction
with Luttinger liquids. Also, the gravitation dressing leads to logarithmic
conformal field theory properties of the four-point function \cite{bilal},
which is also not the case with Luttinger liquids.
Thus gravitationally dressed Fermi liquids are similar, yet distinguishable
from Luttinger liquids.

We note that the generic reasoning associated with 
the gravitational dressing of the renormalization group \cite{kpz, kkp, kawai, bilal} leads us to two universality classes:
a weakly dressed Fermi liquid, in which the gravitational dressing is irrelevant
close to the Fermi surface, and a strongly dressed Fermi liquid, in which
the gravitational dressing is marginally relevant close to the Fermi surface.
The weakly dressed case would be in the universality class of
canonical Fermi liquids, and the strongly dressed case would correspond
to a non-Fermi liquid, namely a natural higher dimensional generalization
of the Luttinger liquid, with specific properties.

Naturally, the anomalous dimensions of Fermi propagators have been
considered previously, in the Marginal Fermi liquid theory, through couplings 
with an effective induced gauge field, in the context of quantum 
critical fixed points, and other proposals regarding the normal state of
high $T_c$ materials \cite{htcreview, nfl, pw, marginal}. Our 
proposal is not unique in
placing emphasis on the anomalous fermionic propagators. Also, the
phenomenology implied by anomalous fermionic propagators, including
non-Fermi liquid behavior, such as a resisitivity linear in temperature $T$ 
\cite{marginal}, is inherent to our proposal as well. However, the proposal that
the anomalous nature of the fermionic two-point function follows from the
concept of a gravitationally dressed Fermi liquid, which is {\it natural}
from the point of view of the renormalization group approach to a
fluctuating Fermi surface, has been, to our knowledge,
first stated in \cite{us}.

\section{Conclusions}

In this note 
we elaborated on the new concept of gravitationally dressed Fermi liquids.
We distinguished between weakly gravitationally dressed Fermi liquids
which fall in the class of the canonical Fermi liquid theory, and strongly
gravitationally dressed Fermi liquids which represent the high dimensional
generalization of Luttinger liquids, with specific features.

In the case of an ordinary Fermi liquid, a one loop calculation of
the beta function does ensure the asymptotic freedom provided the
interaction is perturbatively attractive, as is the case for phonon-electron
interaction, which leads to the strong coupling regime (with bound states
forming) in the infrared \cite{joep}. The resulting wave-function describing the
superconducting state is of a BCS kind \cite{feynman}
\begin{equation}
\Psi_{BCS} \sim \prod_k ( A + B a^*_k {a^*}_{-k}) \Psi_{FL},
\end{equation}
where $A$ and $B$ are the usual variational parameters and $a^{*}$
denote the creation operators of the electron quasiparticle from the Fermi liquid
ground state $\Psi_{FL}$.
The mean field theory capturing the broken symmetry phase is 
given by the Landau-Ginsburg Lagrangian \cite{feynman}
\begin{equation}
S_{LG}=\int d^{D} x L_{LG} (H, \partial H) = \int d^{D}x [\eta^{\mu \nu }\partial _{\mu }H^{\ast }\partial
_{\nu }H-V(|H|^{2})],
\end{equation}%
where the complex order parameter is denoted by $H$ and $V(|H|^{2})$ is an
effective potential (for example, of a quartic type). 

Given the experimental evidence regarding the complexity of
the Fermi surface of  Cu-O or Fe-As compounds \cite{htcreview, nfl}, it is natural to propose that the normal (``strange metal'') state of high $T_c$ superconductors
is described by a gravitationally dressed
Fermi-liquid theory. The gravitationally dressed one-loop beta function would be multiplicatively 
deformed \cite{kkp}. Accordingly, the BCS wave-function would be also
gravitationally dressed. The gravitationally dressed BCS wave-function is
of the same form, except that the creation operators in this case create 
quasiunparticles from the non-Fermi liquid ground state. Note that this discussion is couched in
terms of an effective field theory and therefore does not carry information
about the underlying microscopic mechanism. Thus, the specific
nature of the pairing mechanism of the relevant quasiunparticles and the microscopic relationship of the pairing mechanism with the effective
gravitational description is outside the scope of the present approach. Still, our
approach implies that the usual effective Landau-Ginsburg
theory describing the physics of the quasiunparticle condensate is also gravitationally dressed:
\begin{equation}
S_{LG_{g}}= \int d^{D} x \sqrt{-g} L_{LG} (H, \nabla H) = \int d^{D}x\sqrt{-g}[g^{\mu \nu }\nabla _{\mu }H^{\ast }\nabla
_{\nu }H-V(|H|^{2})].
\end{equation}%
Once again, the weakly dressed Landau-Ginsburg theory corresponds to the usual
superconductors, and the strongly dressed Landau-Ginsburg Lagrangian
applies to high $T_c$ superconductivity.
As pointed
out in our previous paper \cite{us} such a gravitationally dressed Landau-Ginsburg theory
quite naturally leads to a simple geometric mechanism for the high $T_{c}$ of high
temperature superconductors.

\vskip .5cm

\textbf{\Large Acknowledgements}

\vskip .5cm

We would like to thank many colleagues in the condensed matter and
high energy physics communities for their interest in \cite{us}. 
{\small DM} thanks the Abdus Salam International Center for
Theoretical Physics for hospitality and for providing an opportunity to
discuss the ideas presented in this note.
{\small DM} is supported in part by
the U.S. Department of Energy under contract DE-FG05-92ER40677. {\small JJH}
is supported in part by the U.S. Department of Energy under contract DE-FG02-08ER46532.





\vskip 1cm

\begin{thebibliography}{99}
\bibitem{htcreview} E.~W.~Carlson, V.~J.~Emery, S.~A.~Kivelson and D.~Orgad,
``Concepts in High Temperature Superconductivity'', cond-mat/0206217, in
\textit{The Physics of Conventional and Unconventional Superconductors},
Springer, 2008, K.H. Bennemann and J. B. Ketterson, eds.

\bibitem{abh} \textit{Artificial Black Holes}, edited by M. Novello, M.
Visser and G. Volovik; World Scientific, 2002.

\bibitem{volovik} Grigory E. Volovik, \textit{The universe in a helium
droplet}, Oxford University Press, 2003.

\bibitem{us}
D. Minic and J. J. Heremans,
Phys. Rev. {\bf B 78} 214501 (2008) , arXiv:0804.2880.

\bibitem{landau}
L. D. Landau,  Soviet Physics JETP, {\bf 3} 920,
1957; L. D. Landau, Soviet Physics JETP, {\bf 8} 70,
1959.

\bibitem{georgi}
H. Georgi, Phys. Rev. Lett. {\bf 98}, 221601 (2007).

\bibitem{semiholo}
T. Faulkner and J. Polchinski, ``Semi-Holographic Fermi Liquids'',
arXiv:1001.5049.

\bibitem{nfl}
For a recent review and references, consult,
T. Senthil, Phys. Rev. {\bf B 78} (2008) 035103.

\bibitem{joep}  J.~Polchinski, ``Effective Field Theory And The Fermi
Surface'',  arXiv:hep-th/9210046, \textit{1992 TASI lectures}, World Scientific,
J. Harvey and J. Polchinski, eds.  Also, R.~Shankar,  
Rev.\ Mod.\ Phys.\ \textbf{66}, 129 (1994);
G. Benfatto and G. Gallavotti, J. Stat. Phys. 59 (1990) 541;
Phys. Rev B42 (1990) 9967;
R. Shankar, Physica A177 (1991) 530 and original references therein.

\bibitem{flbooks}
P. Nozieres, { \it Theory of Interacting Fermi Systems}, Benjamin, New York, 1964;
G. Baym, C. Pethick, {\it Landau Fermi-Liquid Theory}, Wiley, New York, 1991.


\bibitem{pw} For an insightful review and references consult P.W. Anderson, 
\textit{The theory of superconductivity in the High-$T_C$ Cuprates},
Princeton, 1997 and references therein.

\bibitem{wen}
X.-G. Wen, {\it Quantum Field Theory and Many-Body Systems}, Oxford, 2004,
section 5, and references therein. 



\bibitem{unruh} W.G.~Unruh, Phys. \ Rev. \ Lett. \ \textbf{46} (1981) 1351;
D. Kominis,V. Koulovassilopoulos, Phys. \ Rev. \ D \ \textbf{51} (1995) 282.


\bibitem{gphase}
A. Shapere and F. Wilczek, {\it Geometric Phases in Physics},
World Scientific, 1989.

\bibitem{gqt}
A. Ashtekar and T. A. Schilling,
``Geometrical Formulation of Quantum Mechanics '', arXiv:gr-qc/9706069;
D. Minic and C.-H. Tze, Phys.Rev.{\bf D68}061501 (2003);Phys.Lett.{\bf B581} 111 (2004). 

\bibitem{boson} A.~Luther, \ Phys. \ Rev. \ B \textbf{19}, 320 (1979);
R.~Abhiraman and C.~M.~Sommerfield,  
arXiv:hep-th/9501008;  C.~M.~Sommerfield,  
in Celebration of 60th Birthday Anniversary of Feza Gursey, Apr 1981, Yale
Gursey Sympos.1981. A.~Enciso and A.~P.~Polychronakos,  
Nucl.\ Phys.\ B \textbf{751}, 376 (2006); F.D.M. Haldane, Varenna 1992
lectures and cond-mat/0505529; A. Houghton and J.B. Marston, \ Phys. \ Rev.
\ B \textbf{48}, 7790 (1993); A. H. Castro Neto and E. Fradkin, \ Phys. \
Rev. \ Lett. \textbf{72}, 1393 (1994), and references therein.

\bibitem{kpz} The original reference of gravitational dressing is
V.~G.~Knizhnik, A.~M.~Polyakov and A.~B.~Zamolodchikov,  
Mod.\ Phys.\ Lett.\ A \textbf{3}, 819 (1988); see also, A. Polyakov, \ Mod. \ Phys. \ Lett.
{\bf A2} (1987) 893.

\bibitem{kkp}
I. R. Klebanov, I. I. Kogan and A. M. Polyakov,
\ Phys.\ Rev.\ Lett. {\bf 71} (1993) 3243,
arXiv:hep-th/9309106 and references therein.

\bibitem{kawai}
H. Kawai and M. Ninomiya, \ Nucl. \ Phys. {\bf B336} (1990) 115 and references therein.

\bibitem{bilal}
A. Bilal and I. Kogan, \ Nucl. \ Phys. {\bf B449} (1995) 569 and references therein.

\bibitem{marginal} See for example, C.M.Varma, P.B. Littlewood,
S.Schmitt-Rink, E.Abrahams and A.E.Ruckenstein, \ Phys. \ Rev. \ Lett. 
\textbf{63} 1996 (1989); see also, E.~Kapit and A.~LeClair,
J.Phys. {\bf A42} 025402 (2009),
arXiv:0805.4182 [cond-mat.str-el] and references therein.

\bibitem{feynman} For an insightful review see for example, R.~P.~Feynman, \textit{Statistical
mechanics : a set of lectures}, Reading, Mass.: Addison-Wesley, 1998.
A. A. Abrikosov, L. P. Gorkov, I. E. Dzyaloshinskii, \textit{Quantum Field Theoretical Methods in Statistical Physics}, Dover, 1963;
J. R. Schrieffer, {\it Theory of Superconductivity}, Benjamin/Cummings, Menlo Park, 1964 and references therein.

\end{thebibliography}
\end{document}